 \documentstyle[amssymb,prb,version2,preprint,aps,colordvi]{revtex}
 \begin{document}
 \begin{title}
 A central role for the exchange-correlation hole \\
  in the 2D metal-insulator transition
 \end{title}
\author{D.J.W. Geldart$^{\star,\dagger}$, D. Neilson$^\star$, 
and J.S. Thakur$^\star$}
 \begin{instit}
$^\star$ School of Physics, The University of New South Wales, Sydney 2052 Australia
\end{instit}
 \begin{instit}
$^\dagger$ Department of Physics, Dalhousie University, Halifax, NS,
 Canada B3H3J5      
\end{instit}
 \begin{abstract}
 
The 2D metal-insulator transition can be induced either by decreasing the carrier density or by increasing the spin polarization by applying a magnetic field parallel to the plane. Using experimental results for the shift in critical carrier density in an applied field, we show that the two-electron pair correlation function at short distance ($k_Fr\alt2$) has a universal form along the transition line in the magnetic field-carrier density plane. This result provides direct evidence of the central role of Coulomb repulsion and the exchange-correlation hole in driving the metal-insulator transition.

 {73.20.Dx,71.30.+h,73.40.-c}
 \end{abstract}

 The metal-insulator transition is a problem of great current 
interest \cite{Kravchenko}.  In spite of intense investigation there is still 
no consensus on understanding the basic features of the problem and there is no 
microscopic theory.  It is 
therefore useful to investigate phenomenological aspects of the metal-insulator
transition.  In this paper we point out an interesting 
empirical relationship between the dependence of the 2D pair correlation 
function $g({\mathbf r})$ on a magnetic field parallel to the plane of a two 
dimensional electron liquid 
and the freezing line in the phase diagram of the metal-insulator transition.

It is possible to regard the transition from the metallic state to an 
insulating state as a special case of freezing of the electron liquid.  
Phenomenological criteria in terms of pair correlation functions and 
static structure factors for the freezing of both classical and quantum 
liquids in the absence of impurities 
have been known for some time.  The properties of the static 
structure factor $S({\mathbf k})$ near freezing show common features along the 
freezing line for a wide class of 3D interacting systems, and equivalent 
statements can be made in terms of the correlation function $g({\mathbf 
r})=1+\int{d^3{\mathbf k}}/{(2\pi)^3}\, \text{e}^{i{\mathbf k}.{\mathbf 
r}}\{S({\mathbf k})-1\}$.  Hansen and Verlet \cite{Hansen} established 
for systems of particles interacting through a Lennard-Jones potential 
in three dimensions that the Lindemann ratio along the melting line 
in the temperature-density plane
is practically constant along the melting line.  Hansen and Verlet 
also showed that on approaching the transition from the liquid state, the 
amplitude of the main peak 
in the static structure factor $S({\mathbf k})$ is very nearly constant $\sim 
2.85\pm0.1$ along the crystallization line.  This same 
value has been found for the peak in $S({\mathbf k})$ along the crystallization 
line for both classical one component plasmas and systems of classical hard 
spheres \cite{OCPHS}.  Adler and Wainwright \cite{Adler} obtained analogous 
results for two dimensional systems of hard discs.  Generalizations of these ideas to quantum systems have been given, although the extra role of zero point 
motion makes the effect less clear-cut \cite{Q2D}.
The essential point is that conditions for crystallization of 
the system as a function of external parameters correspond to critical values 
for short distance properties of the pair correlation function, or 
correspondingly the static structure factor.  Consequently, insight into the 
freezing mechanism may be obtained by studying short distance properties 
of pair correlation functions as parameters are varied in the vicinity of the 
transition.  This is clearly simpler than an explicit calculation of the equation of state, specifying full details of the system and its interactions. This can be useful for gaining phenomenological understanding in the 
absence of a microscopic theory.  

This characterization of the freezing line in terms of the strength of short 
distance correlations has been developed for pure systems 
for which the density distribution is uniform.  The electron distribution in the
systems exhibiting a metal-insulator transition is nonuniform but 
since the transition is observed at low levels of impurities, the short 
distance $g({\mathbf r})$ for $k_Fr<2$ should be well approximated by the 
corresponding function for the pure system.  
Diffusion quantum Monte Carlo (DQMC) simulations \cite{Ceperley,Senatore} give us the ground state properties of the 2D system without impurities including $g({\mathbf r})$.  For the pure system the transition from liquid to solid 
occurs at extremely low densities corresponding to $r_s=37\pm5$, where $r_s$ is the average electron spacing in effective Bohr radii.  The preliminary 
quantum Monte Carlo calculations of Chui and Tanatar \cite{Chui} 
which include the effect of charged impurities have suggested that impurities 
may stabilize a solid ground 
state at densities as high as $r_s=10$.  Although a precise value of $r_s$ for 
the transition was not determined, this result does indicate 
that low levels of static defects in the form of a weak quasi-random potential 
may facilitate solidification.

The metal-insulator transition occurs in very high mobility quasi-2D 
semiconductor devices with low carrier densities, $r_s\agt5$, (although 
very recently Altshuler {\it et al}\ \ \cite{Altshuler} have discussed 
data suggesting that the phenomenon may persist up to densities as high 
as $r_s\simeq2$).  At the very low densities of the Wigner 
crystal, $r_s\agt37$, the strong electron correlations generate a 
large area of near-zero electron density surrounding each electron 
\cite{Ceperley}.  This repulsive hard core region forms part of the 
electron's exchange-correlation hole, the complete density profile of 
which is known from the DQMC $g({\mathbf r})$. \cite{Ceperley}  The results 
in Ref.\ \onlinecite{Ceperley} establish that 
the radius of the exchange-correlation hole shrinks as the electron density is 
increased above the solidification density $r_s\simeq37$.  However, the hard 
core effect persists in the electron liquid up to a density corresponding to 
$r_s\simeq7$.  This is a factor some $25$ 
times greater than the density at the Wigner transition.  

The DQMC numerical simulations in Refs.\ \onlinecite{Ceperley,Senatore} show 
that the exchange-correlation hole is stronger for spin 
polarized electrons than for unpolarized electrons.  This is due to the 
additional exchange acting between the increased proportion of parallel 
spin electrons.  In Fig.~1 we see that polarizing the spins at a fixed density 
corresponding to $r_s=9$ significantly expands the relative area of the hard 
core region in $g({\mathbf r})$.  
By $r_s=7$ the region of zero density around each 
electron has disappeared for the unpolarized system, but for the 
polarized system at the same density it is still present.  

We have proposed \cite{TN3} that the reported suppression of the metallic state 
by a parallel magnetic field 
\cite{Hamilton,Hanein}, is directly associated with the expansion of the area of 
the hard core caused by the partial alignment of the spins in the magnetic 
field.  We found that fields $H_\parallel<1$ T destabilize the metallic phase.
The enhancement in the exchange means that the critical impurity density 
$n^c_{i}$ at the transition for the polarized system is smaller than the 
$n^c_{i}$ for the unpolarized system at the same carrier density.  

Hamilton {\it et al} \ \cite{Hamilton} reported that the 
metal-insulator transition boundary in $p$-GaAs is shifted by a 
parallel field $H_\parallel=0.6$ T from hole density 
$p_s=7.5\times10^{10}$ cm$^{-2}$ to $p_s=12.4\times10^{10}$ cm$^{-2}$, 
or correspondingly, from $r_s=9$ to $r_s=7$. This is shown in 
Fig.~2.  The transition boundary gives the critical 
magnetic field $H_c(r_s)$ as a function of carrier density.  Moving along the 
boundary in the direction of increasing density, the value of $H_c$ increases, 
thus increasing the alignment of the electron spins.  

We connect magnetic field and spin polarization by estimating the degree of 
polarization as a function of a 
parallel magnetic field $H_\parallel$ using DQMC data for the pure 
electron system taken from Rapisarda and Senatore \cite{Senatore}.  
In Fig.~3 we show the critical 
$H_c$ needed to fully polarize the electron spins, determined by equating the 
Zeeman energy gain to the energy difference, 
$(g\mu_B/\hbar)H_c(r_s)=[E_p(r_s)-E_u(r_s)]$, where  
$E_p(r_s)$ and $E_u(r_s)$ are the energies per electron for the fully spin 
polarized and unpolarized states, respectively.  We take 
$(g\sigma_z)=1.1$ for GaAs \cite{Daneshvar}.  For fields 
$H_\parallel<H_c(r_s)$ we use a linear relation to determine the degree of 
polarization.  At $r_s=7$ a field $H_\parallel=0.6$ T induces 50\% 
spin polarization.

We emphasize that the pair correlation function in the vicinity of the 
transition is not expected to be significantly affected at short distances by 
low levels of disorder.  For this reason 
the DQMC $g({\mathbf r})$ for the defect free system can be used for 
the short ranged pair correlation function at the transition.  
Our main result is shown in Fig.~4 where we see for $k_Fr\leq2$ that
the $g({\mathbf r})$ on the transition boundary 
remains fixed all the way along the 
observed transition line.  This indicates that for a given sample 
the transition is 
determined by a fixed functional form of the short-range $g({\mathbf r})$, 
and is equivalent to a critical density profile of the exchange-correlation 
hole.  The fixed functional  form for $g({\mathbf r})=g({\mathbf r},H_c(r_s))$ 
suggests a fixed functional form for $S({\mathbf k})$.  This is reminiscent of 
the role of $S({\mathbf k})$ in characterizing the phase boundary in a wide 
range of liquid-``solid'' transitions.  It indicates that the 
metal-insulator transition is driven by 
electron-electron interactions and suggests that the transition is an 
intrinsic many-electron effect and not an artefact of material dependent 
properties.

We conclude that the metal-insulator boundary for the sample as a function of 
spin polarization and carrier density is determined by a fixed 
functional form of the two-electron correlation function at short distances.  
By defining an exchange-correlation hole radius $r_c$ as the value of $r$ at which $g({\mathbf r})$ exceeds a fixed value $<<1$ such as $0.01$, then along the transition line as a function of parallel magnetic field $H$ and carrier density $\rho$ we have,
\begin{equation}
\frac{\partial r_c}{\partial \rho}\delta\rho+\frac{\partial r_c}{\partial H}\delta H = 0\ .
\end{equation}
This means that a knowledge of the dependence of $g({\mathbf r})$ on $\rho$ and $H$ permits us to predict the value of the gradient $\partial H_c/\partial\rho|_{crit-line}$ of the phase boundary. 

 \acknowledgements

This work is supported by an Australian Research Council Grant 
and by the Natural Sciences and Engineering Research Council of Canada.  
We thank Mukunda Das, Alex Hamilton, Michelle Simmons and Lessek \'{S}wierkowski 
for their useful comments.

\figure{Two-electron correlation function $g({\mathbf r})$ taken from Ref.\ 
\onlinecite{Ceperley} at $r_s=9$ for unpolarized system (dotted line) 
and fully polarized system (dashed line), showing the effect of exchange 
enhancement.  
  \label{g1(r)}}
 \ \\
\figure{Plot of phase boundary of metal-insulator transition 
as a function of parallel magnetic field and carrier density $r_s$ 
taken from Ref.\ \onlinecite{Hamilton}.} 
 \ \\
\figure{ Critical parallel magnetic field $H_c$ as a function of $r_s$. 
\hfill}
 \ \\
\figure{Correlation function $g({\mathbf r})$ along the observed 
metal-insulator transition boundary.  Dotted line: $r_s=9$, unpolarized.  
Dash-dot line: $r_s=8$ for $H_\parallel=0.4$ T.  Dashed line: $r_s=7$ for 
$H_\parallel=0.6$ T.  The three curves are essentially identical, indicating a 
universality of the short range correlations at the transition.  
  \label{g2(r)}}

 
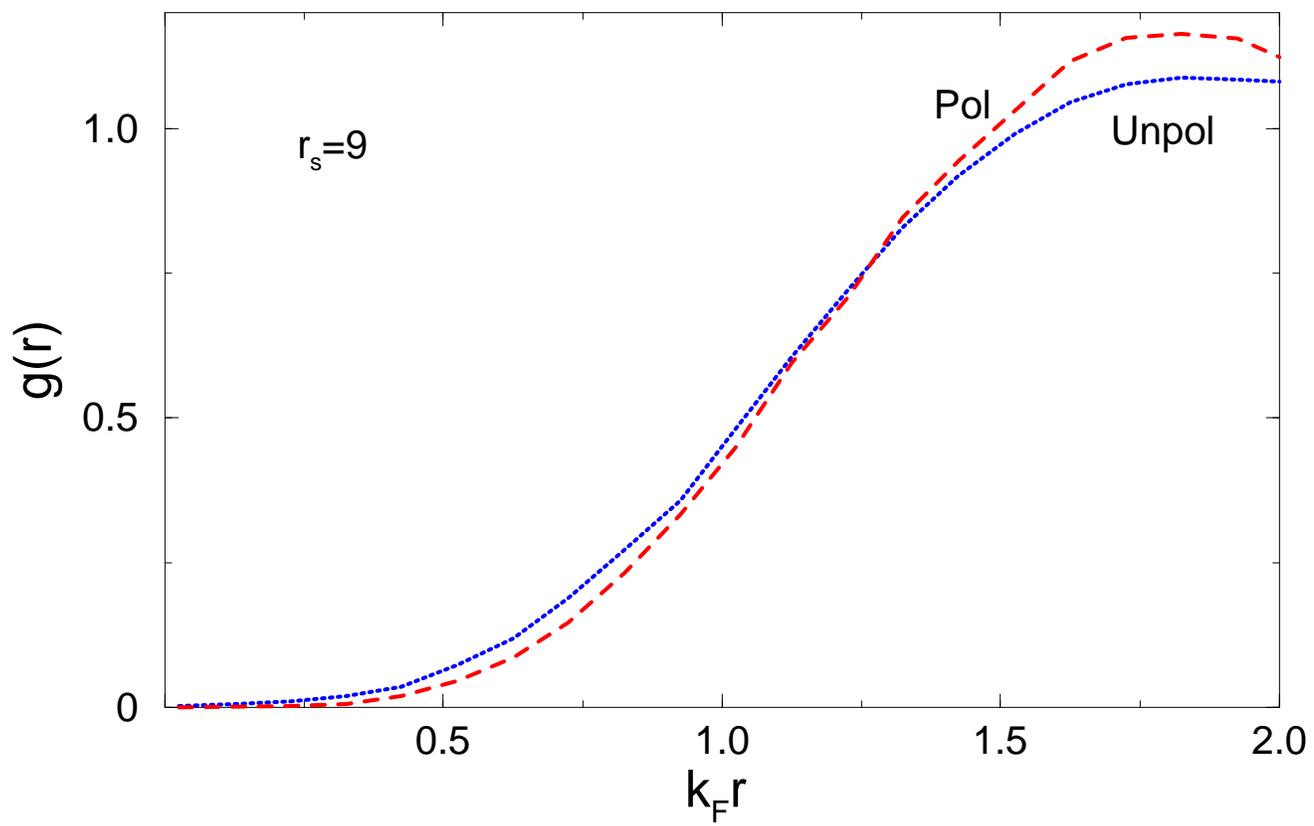
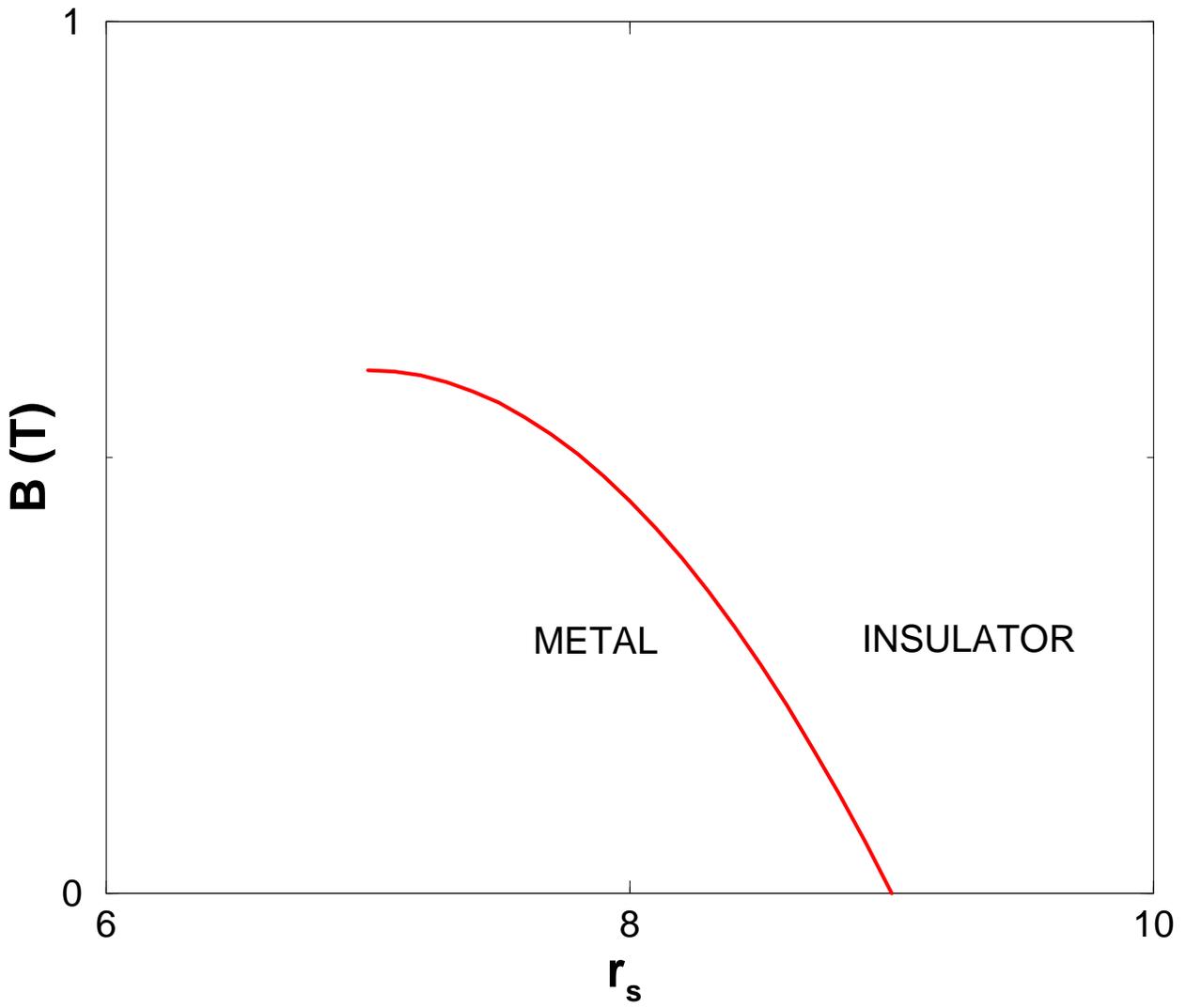
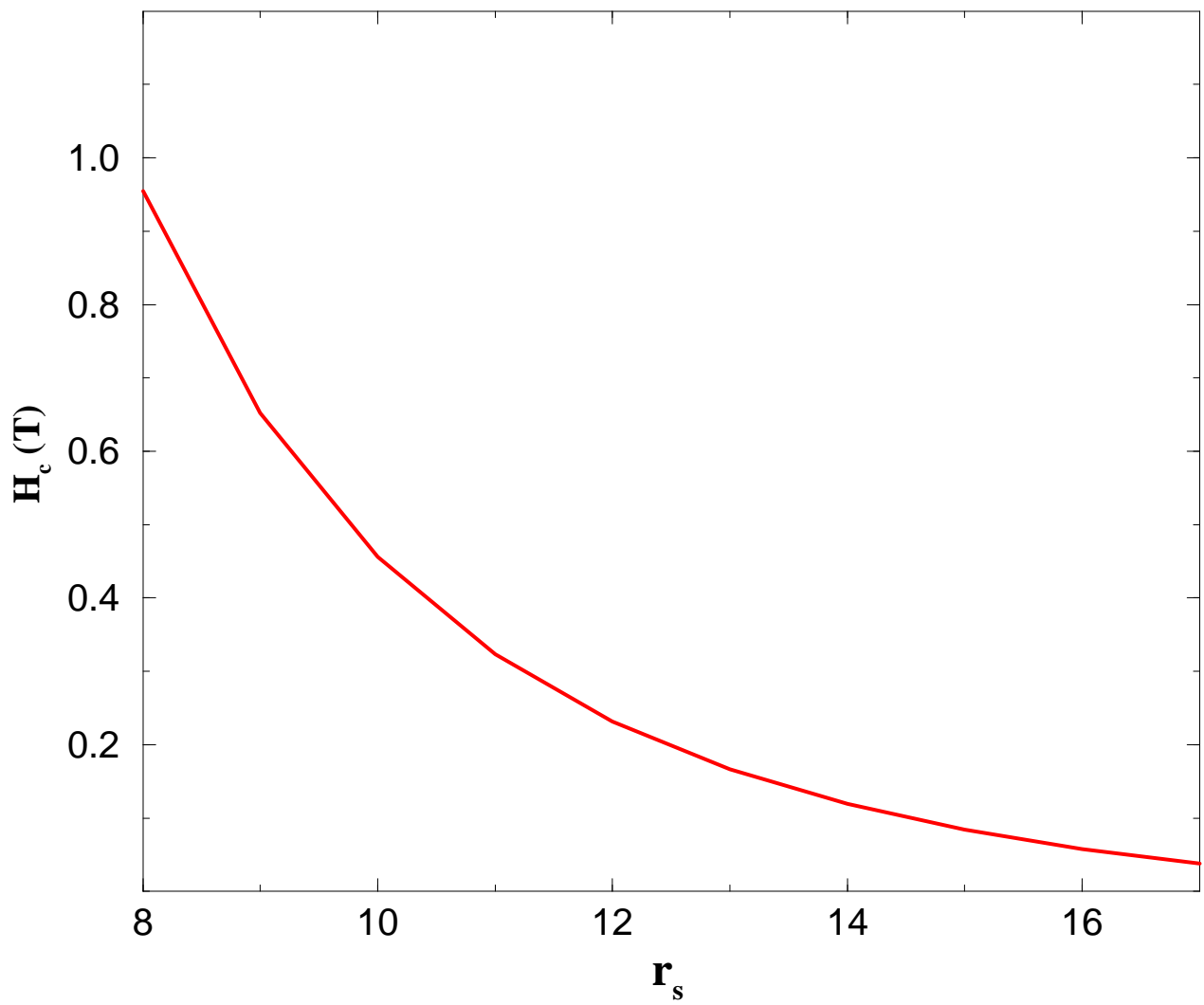
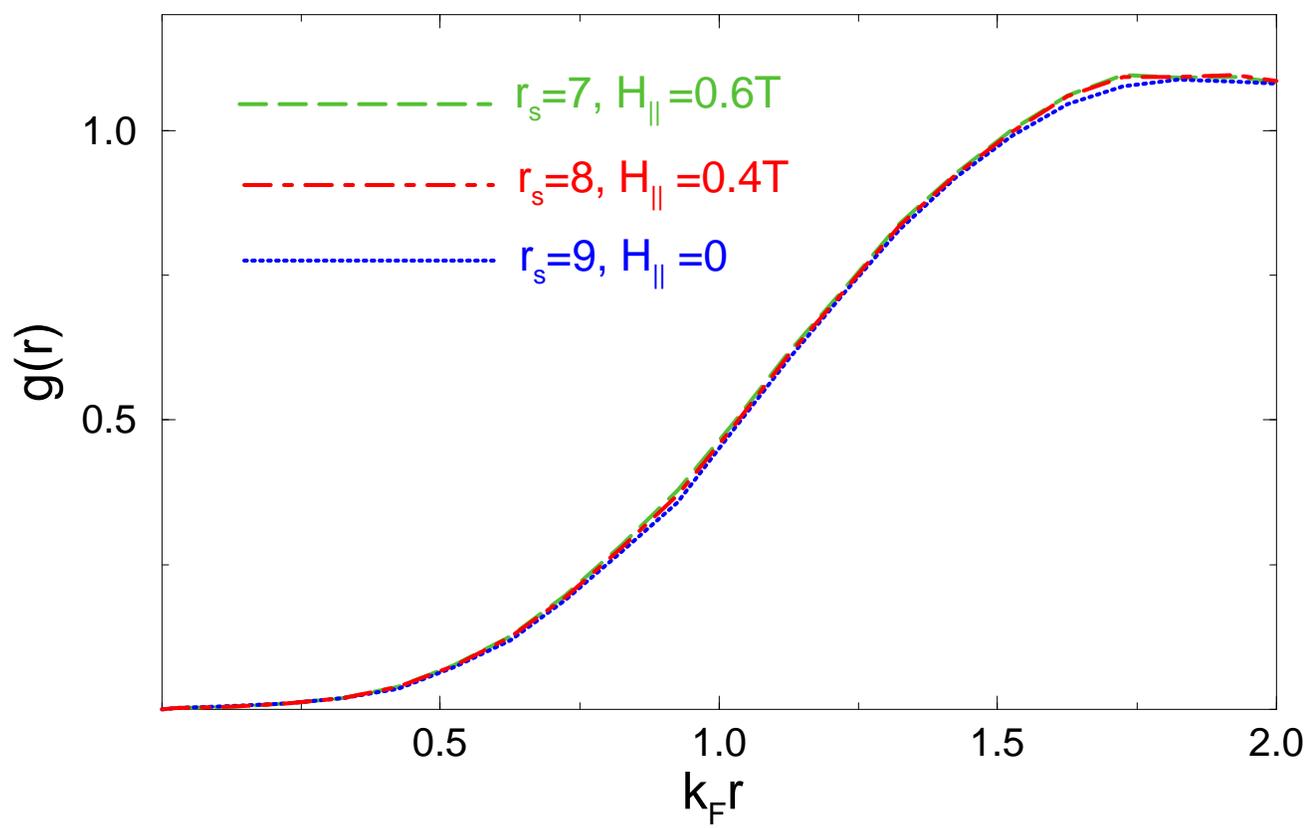
\end{document}